\documentclass[showkeys,showpacs,indentfirst,11pt]{revtex4}
\usepackage{epsfig}
\usepackage{epsf}
\usepackage{bm}
\usepackage{multirow}
\begin{document}

\title{On the structure of the $\pi\pi$ invariant mass spectra of the $\Upsilon(4S)\to\Upsilon(1S,2S)\pi^+\pi^-$ decays}

\author{Feng-Kun Guo$^{1,2,6}$\footnote{Now at Institut f\"ur Kernphysik (Theorie), Forschungszentrum J\"ulich, D-52425 J\"ulich, Germany}}
\email{f.k.guo@fz-juelich.de}
\author{Peng-Nian Shen$^{2,1,4,5}$}
\author{Huan-Ching Chiang$^{3,1}$}
\author{Rong-Gang Ping$^{1,2}$}
\affiliation{\small $^1$Institute of High Energy Physics, Chinese
Academy of Sciences,
P.O.Box 918(4), Beijing 100049, China\\
$^2$CCAST (World Lab.), P.O.Box 8730, Beijing 100080, China\\
$^3$South-West University, Chongqing 400715, China\\
$^4$Institute of Theoretical Physics, Chinese Academy of Sciences, P.O.Box 2735, China\\
$^5$Center of Theoretical Nuclear Physics, National Laboratory of
Heavy Ion Accelerator, Lanzhou 730000, China\\
$^6$Graduate University of Chinese Academy of Sciences, Beijing
100049, China}
\date{\today}

\begin{abstract}
We perform a model-independent analysis for recently reported data
of the $\pi^+\pi^-$ invariant mass spectra in the
$\Upsilon(4S)\to\Upsilon(1S,2S)\pi^+\pi^-$ decays and point out that
there does exist a broad peak below 0.6 GeV in the data of the
$\Upsilon(4S)\to\Upsilon(1S)\pi^+\pi^-$ decay, which is analogous to
that in the $\Upsilon(3S)\to\Upsilon(1S)\pi^+\pi^-$ decay. With the
data of $\Upsilon(4S)$ decays, we further test our model developed
for studying the puzzle in the
$\Upsilon(3S)\to\Upsilon(1S)\pi^+\pi^-$ decay. The result shows that
with such a model, all the $\pi^+\pi^-$ invariant mass spectra of
$\Upsilon(4S)$ decays can be described. We also predict the
$\cos\theta_{\pi}^*$ distributions of
$\Upsilon(4S)\to\Upsilon(1S,2S)\pi^+\pi^-$ decays, which can be used
to justify our model prediction.
\end{abstract}


\maketitle

$\Upsilon(4S)$ is the first bottomonium state above the $B{\bar B}$
threshold. The branching fraction for $\Upsilon(4S)\to B{\bar B}$ is
larger than 96\% \cite{pdg06}. Recently, the $\pi\pi$ invariant mass
spectrum of the $\Upsilon(4S)\to\Upsilon(1S)\pi^+\pi^-$ decay was
measured by the Belle Collaboration \cite{be05,be07} and the BaBar
Collaboration \cite{ba06}, respectively, and the decay of
$\Upsilon(4S)\to\Upsilon(2S)\pi^+\pi^-$ was reported by the BaBar
Collaboration \cite{ba06}. Their results with rather large error
bars are shown in Fig. \ref{fig:up4Sfit}. The branching ratios of
decay modes reported by different collaborations are listed in Table
\ref{tab:Bup4S}.
\begin{table}[htb]
\caption{\label{tab:Bup4S}The branching ratios of the decay modes
$\Upsilon(4S)\to\Upsilon(1S,2S)\pi^+ \pi^-$.}
\begin{center}
\begin{tabular}{|l|c|c|}
\hline
  & $B(\Upsilon(4S)\to\Upsilon(1S)\pi^+ \pi^-) (10^{-4})$ &
  $B(\Upsilon(4S)\to\Upsilon(2S)\pi^+ \pi^-) (10^{-4})$ \\\hline%
CLEO \cite{cl99} & $<1.2$ & $<3.9$ \\\hline%
BaBar \cite{ba06} & $0.90\pm0.15$ & $1.29\pm0.32$ \\\hline%
Belle \cite{be05} & $1.1\pm0.2\pm0.4$ & \\\hline%
Belle \cite{be07} & $1.8\pm0.3\pm0.2$ &\\\hline%
\end{tabular}
\end{center}
\end{table}
Although the branching ratio in Ref. \cite{be05} seems inconsistent
with the newly reported one in Ref. \cite{be07}, it is compatible
with those in Refs. \cite{cl99,ba06}.

An interesting phenomenon in the the
$\Upsilon(3S)\to\Upsilon(1S)\pi^+\pi^-$ decay is that the
$\pi^+\pi^-$ invariant mass spectrum has a double peak structure,
namely a broad peak shows up below 0.6 GeV \cite{cleo}. In past
years, much attention has been attracted to such a phenomenon
\cite{gs05,lt88,mox,up31,ck93,absz,mu97}. Now, the reported
$\pi^+\pi^-$ invariant mass spectrum of the
$\Upsilon(4S)\to\Upsilon(2S)\pi^+ \pi^-$ decay also exhibits a
distinguishable double peak structure, but in the
$\Upsilon(4S)\to\Upsilon(1S)\pi^+ \pi^-$ decay, the lower structure
seems not evident \cite{cl98,be05,ba06}. Based on this observation,
Vogel suspected that the dipion transitions between two
$\Upsilon(nS)$ states with $\Delta n=2$ are special \cite{vo06}.

In order to explain the double peak structure and the angular
distribution in the $\Upsilon(3S)\to\Upsilon(1S)\pi^+\pi^-$ decay,
we proposed an additional sequential process in our model. In such a
process, an intermediate $b{\bar b}q{\bar q}$ state with $J^P=1^+$
and $I=1$ (called $X$) is suggested. By including this sequential
process and considering the $S$ wave $\pi\pi$ final state
interaction (FSI), we well-described the
$\Upsilon(nS)\to\Upsilon(mS)\pi^+\pi^-$ ($n=2,3$, $m=1,2$ and $n>m$)
data available in a systematical way \cite{gs05}. Obviously, the
newly reported data will provide us an opportunity to justify our
model and check Vogel's assertion.

In this paper, we perform a model-independent analysis for all the
$\pi^+\pi^-$ invariant mass spectrum data in the
$\Upsilon(4S)\to\Upsilon(1S,2S)\pi^+ \pi^-$ transitions. Although
the Belle Collaboration claimed that the data in Ref. \cite{be05}
should be superseded by those in Ref. \cite{be07}, we still use all
these data to justify our model proposed in Ref. \cite{gs05}, and
consequently, predict the angular distribution for further
confirmation in future.

Before processing detailed calculations, we analyze the $\pi^+\pi^-$
invariant mass spectra of the $\Upsilon$ $\pi\pi$ transitions,
qualitatively.

1) The isoscalar character of bottomonia limits the isospin of the
dipion system to be $I=0$. Because both $\Upsilon(4S)$ and
$\Upsilon(1S)$ are the vector state, the $\pi^+\pi^-$ system favors
the $S$ wave than the higher ones. A mass difference of 1.12 GeV
between $\Upsilon(4S)$ and $\Upsilon(1S)$ gives the upper limit of
the physical $M_{\pi^+\pi^-}$ region. In this region, the
well-established isoscalar-scalar meson of $f_0(980)$ would couple
to $\pi\pi$ strongly, which causes its width of about 40-100 MeV
\cite{pdg06}. Therefore, there might be a narrow peak or dip in the
$\pi^+\pi^-$ invariant mass spectrum of the
$\Upsilon(4S)\to\Upsilon(1S)\pi^+\pi^-$ decay. However, if the bin
in the experimental data is wide, as shown in all the reported data
sets, the narrow state of $f_0(980)$ might not show up explicitly.

\begin{figure}[htbp]
\begin{center}\vspace*{0.cm}
\includegraphics[width=0.5\textwidth]{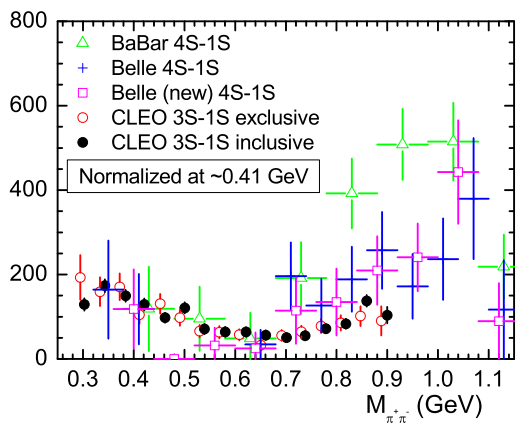}%
\vglue-0.5cm\caption{\label{fig:up4Snorm} The $\pi^+ \pi^-$
invariant mass spectra of the $\Upsilon(3S,4S)\to\Upsilon(1S)\pi^+
\pi^-$ decays without the phase space factor.}
\end{center}
\end{figure}

2) In comparison with the measured invariant mass spectrum, namely
the differential width, the squared decay amplitude ($|{\cal M}|^2$)
reflects the decay dynamics more directly. Among low partial waves
between two pions, only $S$- and $D$-waves can contribute to the
decay amplitude. Because the contribution from the $D$-wave is
suppressed by a factor of 2/45 due to the integration of the 2nd
order Legendre polynomial, the squared decay amplitude can be
approximated by the invariant mass spectrum divided by the
integrated phase space. The $\pi^+ \pi^-$ invariant mass spectra of
the $\Upsilon(3S,4S)\to\Upsilon(1S)\pi^+ \pi^-$ decays without the
phase space factor are shown in Fig. \ref{fig:up4Snorm}. For
comparison, we normalize the magnitudes around $m_{\pi^+\pi^-}=0.41$
GeV in all the data sets from Refs. \cite{be05,ba06,be07} for the
$\Upsilon(4S)\to\Upsilon(1S)\pi^+\pi^-$ decay and from Ref.
\cite{cleo} for the $\Upsilon(3S)\to\Upsilon(1S)\pi^+\pi^-$ decay to
the values close to each other. From this figure, one sees that the
tendency of the data in the $\Upsilon(4S)\to\Upsilon(1S)\pi^+\pi^-$
decay, especially in the region lower than $0.65$ GeV, is analogous
to that in the $\Upsilon(3S)\to\Upsilon(1S)\pi^+\pi^-$ decay. This
indicates that in the $\pi\pi$ invariant mass spectrum of the
$\Upsilon(4S)\to\Upsilon(1S)\pi^+\pi^-$ decay, a broad peak does
exist in the lower $M_{\pi\pi}$ region, which implies that the above
mentioned $\Delta n=2$ rule might not be true.

\begin{figure}[htb]
\begin{center}
{\epsfysize=2cm \epsffile{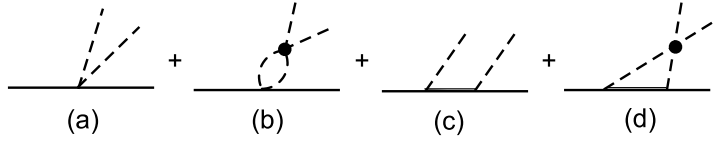}}%
\vglue -0.1cm\caption{\label{fig2}The decay mechanism of
$\Upsilon(nS)\to\Upsilon(mS)\pi^+\pi^-$. Solid external lines
represent $\Upsilon(nS)$, dashed lines represent pions, and solid
intermediate lines represent $X$.}
\end{center}
\end{figure}

\begin{figure}[htbp]
\begin{center}\vspace*{0.0cm}
\includegraphics[width=0.5\textwidth]{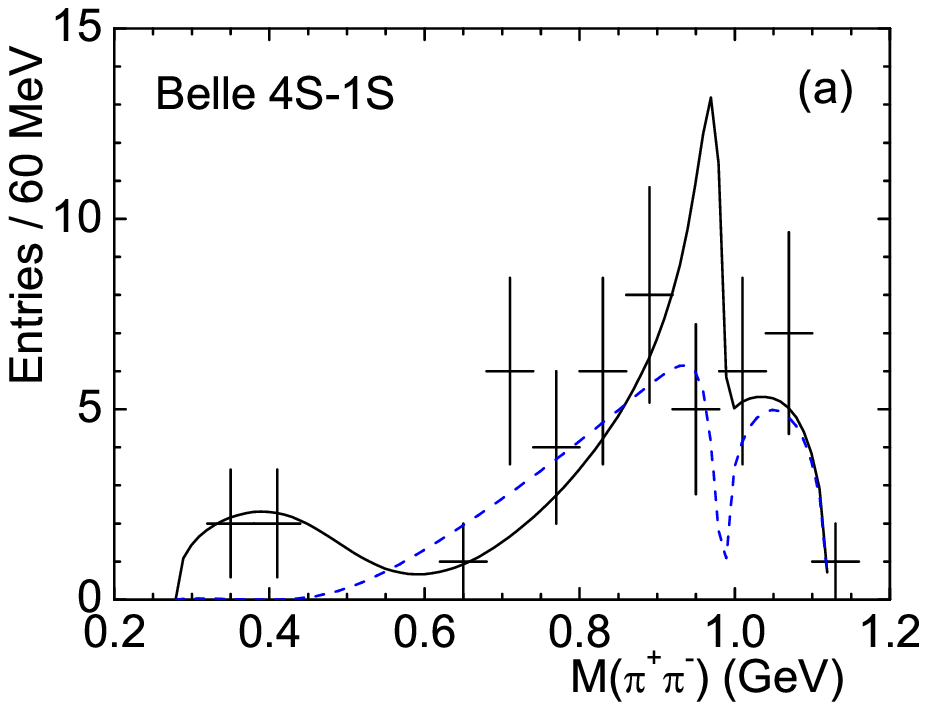}%
\includegraphics[width=0.5\textwidth]{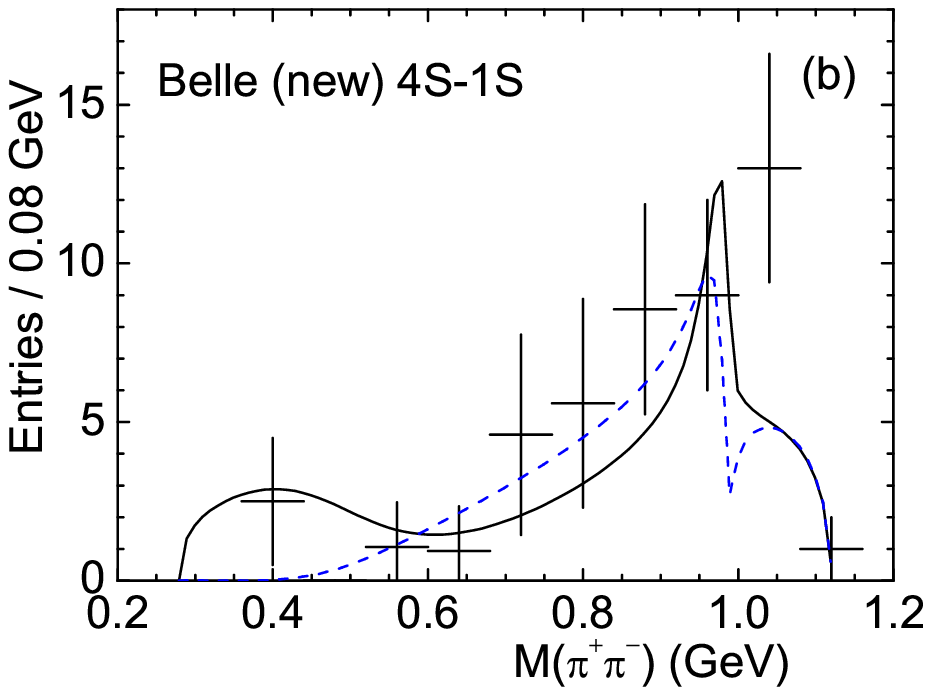}\\\vglue-0.5cm%
\includegraphics[width=0.5\textwidth]{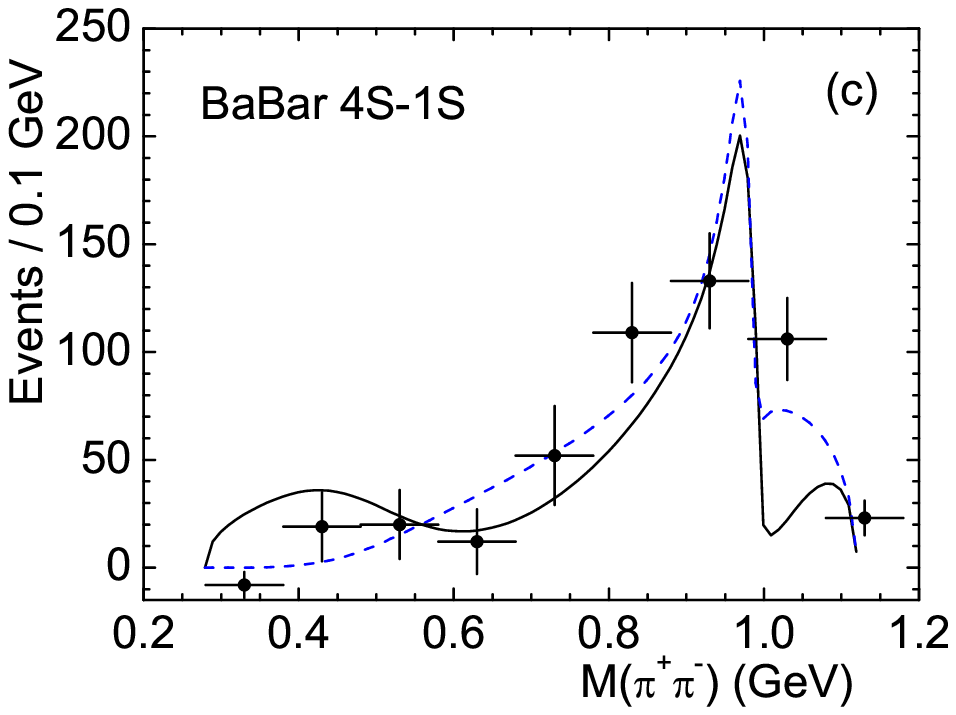}%
\includegraphics[width=0.5\textwidth]{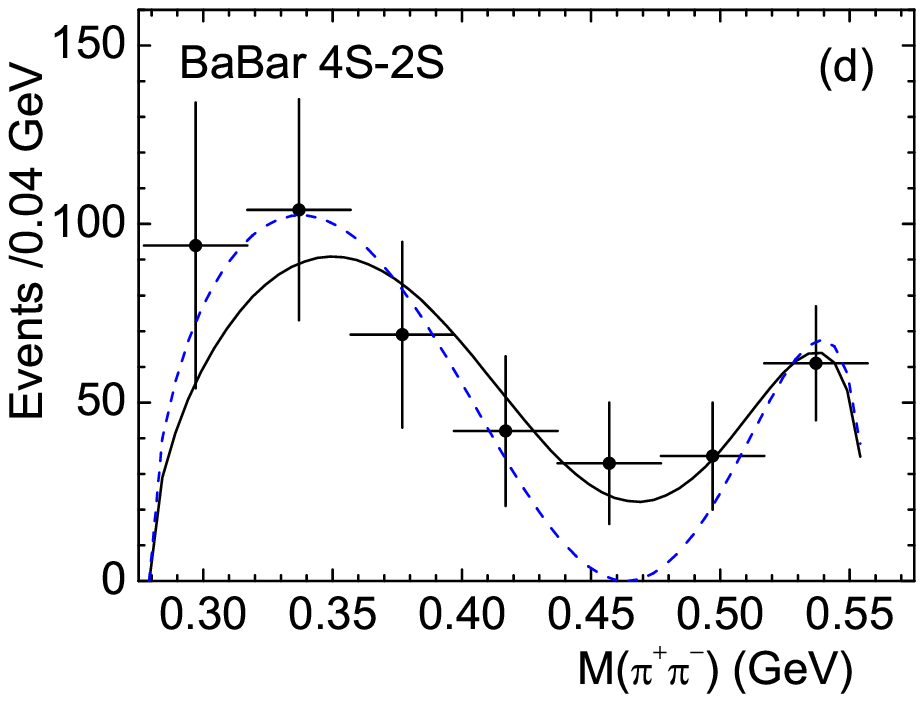}%
\vglue -0.5cm\caption{\label{fig:up4Sfit}The $\pi^+\pi^-$ invariant
mass spectra for the decays $\Upsilon(4S)\to\Upsilon(1S,2S)\pi^+
\pi^-$.  The solid and dashed curves denote the results with and
without $X$, respectively.}
\end{center}
\end{figure}

Now, we use our proposed model \cite{gs05} to study
$\Upsilon(4S)\to\Upsilon(1S,2S)\pi^+\pi^-$ decays. The decay
mechanism of $\Upsilon(nS)\to\Upsilon(mS)\pi^+\pi^-$ is shown in
Fig. \ref{fig2}, where (a) and (c) represent the contact diagram and
the diagram in tree level with $X$, respectively, and (b) and (d)
denote the corresponding diagrams with the $S$-wave $\pi\pi$ FSI.

By incorporating appropriately the chiral expansion with the heavy
quark expansion to the lowest order, the amplitude of the contact
term in the heavy quarkonium $\pi^+ \pi^-$ transition can be written
as \cite{bc75,ck93,mu97}
\begin{equation}
\label{eq:mannel} V_0=-\frac{4}{f^{2}_\pi} \left[(g_1p_1\cdot
p_2+g_{2}p_1^0p_2^0 +g_3m_\pi^2)\varepsilon^*\cdot\varepsilon{'} +
g_4(p_{1\mu}p_{2\nu}+p_{1\nu}p_{2\mu})
\varepsilon^{*\mu}\varepsilon'^{\nu}\right],
\end{equation}
where $f_\pi=92.4$ MeV is the pion decay constant, $g_i~(i=1,2,3,4)$
are coupling constants, $p_1$ and $p_2$ represent the 4-momenta of
$\pi^+$ and $\pi^-$, respectively, $p_i^0~(i=1,2)$ denote the
energies of $\pi^{\pm}$ in the lab frame, and $\varepsilon$ and
$\varepsilon{'}$ describe the polarization vectors of the heavy
quarkonia, respectively. It was shown in Refs. \cite{ck93,mu97} that
the $g_4$-term in Eq.~(\ref{eq:mannel}) can be ignored in the view
point of QCD multipole expansion. Then, Eq.~(\ref{eq:mannel}) can be
reduced to
\begin{eqnarray}
V_{0}=-\frac{4}{f^{2}_\pi}(g_1p_1\cdot p_2+g_{2}p_1^0p_2^0
+g_3m_\pi^2)\varepsilon^*\cdot\varepsilon{'}.
\end{eqnarray}
Note that part of $D$-wave components still exist in the $g_2$ term
\cite{gs05}
\begin{eqnarray}
p_1^0p_2^0=\frac{1}{1-{\bm \beta}^2}\left[\left(p_1^{0*2}-
\frac{{\bm \beta}^2{\mathbf p}_1^{*2}%
}{3}\right)P_0(\cos\theta_{\pi}^*)-2{\bm \beta}^2{\mathbf
p}_1^{*2}P_2(\cos\theta_{\pi}^*)\right],
\end{eqnarray}
where ${{\bm\beta}}$ is the velocity of the $\pi\pi$ system in the
rest frame of the initial particle, ${p_1^*}_\mu=(p_1^{0*}, {\mathbf
p}_1^*$) and ${p_2^*}_\mu=(p_2^{0*},{\mathbf p}_2^*$) represent the
four-momenta of $\pi^+$ and $\pi^-$ in the c.m. frame of the
$\pi\pi$ system, respectively. $P_0(\cos\theta_{\pi}^*)=1$ and
$P_2(\cos\theta_{\pi}^*)=(\cos^2\theta_{\pi}^*-1/3)/2$ are the
Legendre functions of the 0-th order and the 2-nd order,
respectively.

In our model, we stressed that an additional sequential process
$\Upsilon(nS)\to\pi X\to\Upsilon(mS)\pi\pi$ should also contribute
to the decay. Taking a simple $S$-wave coupling for the vertex
$\Upsilon(nS)\to\pi X$, the decay amplitude of the sequential
process in Fig. \ref{fig2} (c) can be written as
\begin{equation}
 \label{eq:xtree}
V_X^{tree}=g_{nm} \epsilon'_{\mu}\epsilon^{*}_{\nu}\left(\frac{%
-g^{\mu\nu}+p_{X^+}^{\mu}p_{X^+}^{\nu}/m_X^2}{p_{X^+}^2-m_X^2+i m_X\Gamma_X}+%
\frac{-g^{\mu\nu}+p_{X^-}^{\mu}p_{X^-}^{\nu}/m_X^2}{p_{X^-}^2-m_X^2+i
m_X\Gamma_X}\right),
\end{equation}
where $g_{nm}$ is the effective coupling constant among
$\Upsilon(mS)$, $\Upsilon(nS)$, $\pi^+$ and $\pi^-$ via an
intermediate state $X$, and $p_{X^{\pm}}$ denote the momenta of the
charged intermediate states of $X^{\pm}$, respectively. In Ref.
\cite{gs05}, we showed that the $X$ state is necessary in getting a
global fit to all the $\Upsilon(nS)\to\Upsilon(mS)\pi^+\pi^-$
$(n=2,3,~m=1,2,~n>m)$ data. Especially in the
$\Upsilon(3S)\to\Upsilon(1S)\pi^+\pi^-$ decay, the contribution of
$X$ is very important for both the dipion invariant mass spectrum
and the angular distribution. The predicted mass and width of the
$X$ state are $m_X=10.05$ GeV and $\Gamma_X=0.688$ GeV, respectively
\cite{gs05}. It should be mentioned that the predicted $X$ state is
supported by later theoretical calculations with the constituent
quark model \cite{fe06} and the QCD sum rules \cite{mn07},
respectively.

In Ref. \cite{gs05}, it was also shown that the $S$-wave $\pi\pi$
FSI is very important to reproduce the double peak structure in the
$\pi^+\pi^-$ invariant mass spectrum of the
$\Upsilon(3S)\to$$\Upsilon(1S)\pi^+\pi^-$ decay. Such an importance
has further been presented in the study of the heavy quarkonium
chromo-polarizability \cite{gs06} that describes the interaction of
heavy quarkonia with soft gluons \cite{vo04}. The $S$-wave $\pi\pi$
FSI can properly be treated by the so-called coupled-channel chiral
unitary approach \cite{oo97}. In this approach, the $\pi\pi$
$S$-wave phase shifts can be well-described with barely one
parameter, 3-momentum cut-off $q_{max}=1.03$ GeV, and the low lying
scalar mesons ($\sigma$, $f_0(980)$, $a_0(980)$ and $\kappa$) can
dynamically be generated with reasonable masses and widths
\cite{oo97,gp05}. In the $S$-wave isoscalar sector, the $\pi\pi$ and
$K{\bar K}$ channels are taken into account (for detailed
information, refer to Ref. \cite{oo97}). By using the phase
convention $|\pi^+\rangle=-|1,1\rangle$ and the normalization
$\langle\pi^+\pi^-|\pi^+\pi^-\rangle =
\langle\pi^-\pi^+|\pi^-\pi^+\rangle =
\langle\pi^0\pi^0|\pi^0\pi^0\rangle = 2$ (considering the fact that
$\pi^+,\pi^-$ and $\pi^0$ are in the same isospin multiplet), the
full amplitude of the
$\pi^+\pi^-+\pi^-\pi^++\pi^0\pi^0\to\pi^+\pi^-$ process can be
denoted by $2t^{I=0}_{\pi\pi,\pi\pi}$ with $t^{I=0}_{\pi\pi,\pi\pi}$
being the full coupled-channel amplitude of the $I=0$ $S$-wave
$\pi\pi\to \pi\pi$ process. In the
$\Upsilon(4S)\to\Upsilon(1S)\pi^+\pi^-$ decay, the large phase space
allows the $K{\bar K}$ channel to be on shell. Thus, the full
amplitude of the $K^+K^-+K^0{\bar K}^0\to\pi^+\pi^-$ process can
similarly be described by $2t^{I=0}_{K{\bar K},\pi\pi}/\sqrt{3}$
with the phase convention $|{\bar K}^0=-|1/2,1/2\rangle$ and the
amplitude $t^{I=0}_{K{\bar K},\pi\pi}$ which denotes the full
coupled-channel amplitude of the $I=0$ $S$-wave $K{\bar K}\to
\pi\pi$ process. To factorize the FSI from the contact term in Fig.
\ref{fig2} (b), the on-shell approximation is adopted as usual. The
off-shell effects can be absorbed into phenomenological coupling
constants. In this approximation, both the $\pi\pi$ and $K{\bar K}$
loops shown in Fig. \ref{fig2} (b) are allowed in the
$\Upsilon(4S)\to\Upsilon(1S)\pi^+\pi^-$ decay, but only the $\pi\pi$
loop is permitted in the $\Upsilon(4S)\to\Upsilon(2S)\pi^+\pi^-$
decay. Then, the total transition amplitude for
$\Upsilon(4S)\to\Upsilon(1S)\pi^+\pi^-$ can be given by
\begin{equation}
\label{eq:total} t=V_0 + V_{0S}\cdot G_{11}\cdot
2t^{I=0}_{\pi\pi,\pi\pi} + V_{0S}(m_K)\cdot G_{22}\cdot
\frac{2}{\sqrt{3}}t^{I=0}_{K{\bar K},\pi\pi} + V_X^{tree} +
g_{41}\epsilon^{'}_{\mu}\epsilon^{*}_{\nu} G_X^{\mu\nu}\cdot
2t^{I=0}_{\pi\pi,\pi\pi},
\end{equation}
and the transition amplitude for
$\Upsilon(4S)\to\Upsilon(2S)\pi^+\pi^-$ by
\begin{equation}
\label{eq:total} t=V_0 + V_{0S}\cdot G_{11}\cdot
2t^{I=0}_{\pi\pi,\pi\pi} + V_X^{tree} +
g_{42}\epsilon^{'}_{\mu}\epsilon^{*}_{\nu} G_X^{\mu\nu}\cdot
2t^{I=0}_{\pi\pi,\pi\pi},
\end{equation}
where $V_{0S}$ is the $S$-wave projection of $V_0$, and
$V_{0S}(m_K)$ is the one for the process with two kaons. $G_{ii}$
and $G_X^{\mu\nu}$ are the two-meson loop integral ($i=1$ for the
$\pi\pi$ loop, and $i=2$ for the $K{\bar K}$ loop) and the
three-meson loop integral, respectively.
\begin{eqnarray}
\label{eq:2loop}
G_{ii}=i\int\frac{d^4q}{(2\pi)^4}\frac{1}{q^2-m_i^2+i \varepsilon} \frac{1}{%
(p^{^{\prime}}-p-q)^2-m_i^2+i\varepsilon},
\end{eqnarray}
\begin{eqnarray}
G_X^{\mu\nu}=i\int\frac{d^4q}{(2\pi)^4} \frac{-g^{\mu\nu}+p_{X}^{\mu}p_{X}^{%
\nu}/m_X^2}{p_X^2-m_X^2+i\varepsilon} \frac{1}{q^2-m_{\pi}^2+i
\varepsilon}
\frac{1}{(p^{^{\prime}}-p-q)^2-m_{\pi}^2+i\varepsilon},
\end{eqnarray}
where $p'$ and $p$ represent the momenta of $\Upsilon(4S)$ and
$\Upsilon(1S)$, respectively. The loops are calculated with a
cut-off momentum $q_{max}=1.03$ GeV which was used in explaining the
data of the $\pi\pi$ $S$-wave scattering \cite{oo97}. Thus, the FSI
of $\pi\pi$ used here is consistent with the interaction of $\pi\pi$
employed in fitting the $\pi\pi$ scattering data. It should be
mentioned that in calculating the transition amplitude for the
diagram in Fig. \ref{fig2}(d), we do not consider the intermediate
$X_s(b{\bar b}s{\bar s})$ state, the strange cousin of $X$, due to
its negligible contribution (refer to Ref. \cite{gs05}). In this
way, no additional parameter is introduced in the calculation.

As has been pointed out in Ref. \cite{gs05}, to describe all the
bottomonium dipion transition data, including the newly reported
data, self-consistently, the $g_2$ value is fixed to be
$g_2/g_1=-0.23$. Then, the data of the $\pi^+\pi^-$ invariant mass
spectra in the $\Upsilon(4S)\to\Upsilon(1S,2S)\pi^+\pi^-$ decays
\cite{be05,ba06,be07} are fitted by three parameters, $g_1$,
$g_3/g_1$ and $g_{nm}$, where $m=1$ and $m=2$ represent the final
states of $\Upsilon(1S)$ and $\Upsilon(2S)$, respectively. Namely,
in terms of the function minimization and error analysis package
(MINUIT) \cite{mint}, these parameters can be determined by a least
square fit. The resultant parameters with relevant errors are listed
in Table \ref{tab:up4Spara}. In order to demonstrate the necessity
of the suggested intermediate state $X$, the data fitting without
$X$ is also performed, and the relevant parameters are presented in
Table \ref{tab:up4Spara} too. Moreover, the central values of the
experimental branching widths,
$\Gamma(\Upsilon(4S)\to\Upsilon(1S)\pi^+\pi^-)=2.2$ keV \cite{be05},
3.7 keV \cite{be07}, 1.8 keV \cite{ba06} and
$\Gamma(\Upsilon(4S)\to\Upsilon(2S)\pi^+\pi^-)=2.7$ keV \cite{ba06},
are used in determining the physical values of coupling constants as
well. It should be noted that due to $g_{nm}=g_{nX}g_{mX}$, the
values of $g_{nm}$ with different $n$ and $m$ are not fully
independent. Using the values of $g_{41}$ and $g_{42}$ given in
Table \ref{tab:up4Spara} and the value of $g_{31}$ given in Ref.
\cite{gs05}, one can deduce $g_{32}=7.36$ GeV$^2$ which is different
from the value of $-0.00418$ GeV$^2$ given in Ref. \cite{gs05}.
Fortunately, because the contribution from the sequential process is
not important in the $\Upsilon(3S)\to\Upsilon(2S)\pi^+ \pi^-$ decay,
such a change does not affect the description of the data.

\begin{figure}[htbp]
\begin{center}\vspace*{0.cm}
\begin{center} 
\hglue0.3cm\includegraphics[width=0.5\textwidth]{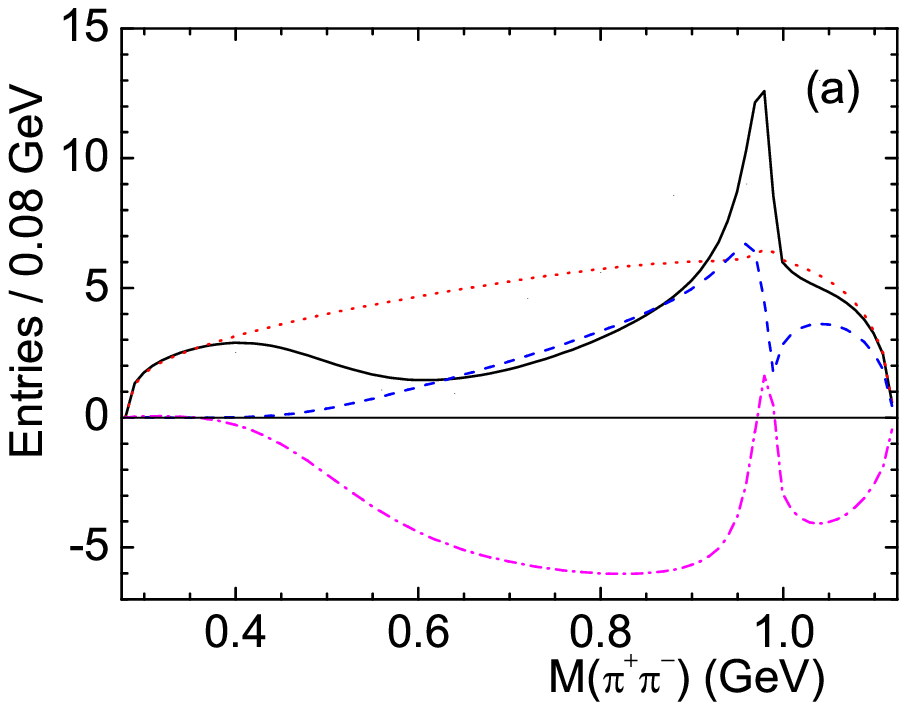}
\hglue-0.6cm\includegraphics[width=0.5\textwidth]{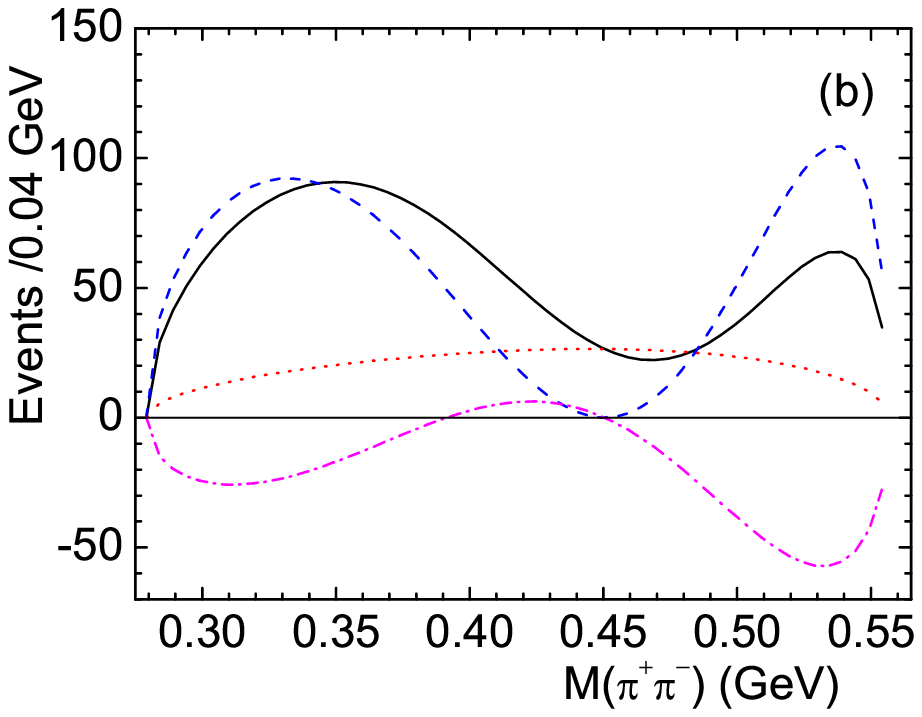}
\vglue-0.5cm{\caption{\label{fig:up4Smana} The contributions from
different terms to the $\pi^+\pi^-$ invariant mass spectra. (a)
$\Upsilon(4S)\to\Upsilon(1S)\pi^+\pi^-$, (b)
$\Upsilon(4S)\to\Upsilon(2S)\pi^+\pi^-$.}}
\end{center}
\end{center}
\end{figure}

The calculated $\pi^+\pi^-$ invariant mass spectra of the
$\Upsilon(4S)\to\Upsilon(1S,2S)\pi^+\pi^-$ decays are plotted in
Fig. \ref{fig:up4Sfit}, where the solid and dashed curves denote the
results with and without $X$, respectively. In the case with $X$, a
clear bump exists in the low-energy region of the $\pi^+\pi^-$
invariant mass spectra in both the
$\Upsilon(4S)\to\Upsilon(1S)\pi^+\pi^-$ and the
$\Upsilon(4S)\to\Upsilon(2S)\pi^+\pi^-$ decays, while in the case
without $X$, no such a structure appears in the
$\Upsilon(4S)\to\Upsilon(1S)\pi^+\pi^-$ decay. It seems that the
newly reported $\Upsilon(4S)$ $\pi\pi$ transition data support our
model in which the sequential process with $X$ is influential.
Moreover, in the $\pi^+\pi^-$ invariant mass spectrum of the
$\Upsilon(4S)\to\Upsilon(1S)\pi^+\pi^-$ decay, our model gives a
narrow structure at the place just below 1 GeV. In addition, in the
$\Upsilon(4S)\to\Upsilon(2S)\pi^+\pi^-$ decay, the calculated
results for Fig. \ref{fig:up4Sfit}(d) show a better description of
the data if the sequential process with $X$ is also considered.
These results can be understood by analyzing the contributions of
different terms. For simplicity, in the case of
$\Upsilon(4S)\to\Upsilon(1S)\pi^+ \pi^-$, we only focus our
attention on the data set in Ref. \cite{be07}. The contributions
from different terms are plotted in Fig. \ref{fig:up4Smana}, where
the solid curve denotes the total result, and the dashed, dotted and
dash-dotted curves represent the contributions from the terms
without $X$, with $X$ only, and the interference term, respectively.
It is seen that the narrow peak close to 1 GeV comes from the
$S$-wave coupled-channel $\pi\pi$ FSI. Recalling that the
isoscalar-scalar meson $f_0(980)$ can be generated dynamically in
the coupled-channel chiral unitary approach \cite{oo97}, one can
attribute the origin of the narrow peak to the effect of $f_0(980)$.
However, due to the limited resolving power of the present data,
this narrow peak might not be observed. Higher statistical data
should be called. If this peak still cannot be found in the future
higher statistical data, one should not be surprised, because it
might be canceled by other stuff. An analogue is that in studying
$J/\psi\to\omega\pi^+\pi^-$ decay in the coupled-channel chiral
unitary approach \cite{rp04}, due to the contribution of the
intermediate state $b_1(1235)$, the peak of $f_0(980)$ cannot be
observed \cite{dm2,bes}. Similar to the situation in the
$\Upsilon(3S)\to\Upsilon(1S)\pi^+\pi^-$ decay \cite{gs05}, the
contribution from the sequential process plays a dominant role in
the $\Upsilon(4S)\to\Upsilon(1S)\pi^+\pi^-$ decay. But, in the
$\Upsilon(4S)\to\Upsilon(2S)\pi^+\pi^-$ decay, the dominant
contribution comes from the terms without $X$. Furthermore, in both
$\Upsilon(4S)$ decays, the contribution from the interference term
is important. Anyway, in order to justify the heavy quarkonium
$\pi\pi$ transition model further, the data with higher statistics
should be requested.

\begin{table}[hbtp]
\begin{center}
\caption{\label{tab:up4Spara} Resultant parameters in fitting the
$\pi^+\pi^-$ invariant mass spectra of the
$\Upsilon(4S)\to\Upsilon(1S,2S)\pi^+\pi^-$ decays.}
\begin{ruledtabular}
\begin{tabular}{cllll}
Decay                & Data source        & ~$g_1$                         & ~~$g_3/g_1$     &      $g_{nm}$ (GeV$^2$) \\\hline%
\multirow{6}*{$\Upsilon(4S)\to\Upsilon(1S)\pi^+\pi^-$}
                       & Belle \cite{be05} & $(6.00\pm1.43)\times10^{-3}$   & ~~$0.69\pm0.63$ & $5.28\pm1.20$      \\%
                       & Belle \cite{be05} & $(6.51\pm1.18)\times10^{-3}$   & ~~$0.44\pm0.48$ & 0 (fixed)          \\%
                       & Belle \cite{be07} & $(5.56\pm4.48)\times10^{-3}$   & ~~$0.94\pm1.74$ & $7.60\pm1.19$      \\%
                       & Belle \cite{be07} & $(6.76\pm1.46)\times10^{-3}$   & ~~$1.04\pm0.52$ & 0 (fixed)          \\%
                       & BaBar \cite{ba06} & $(7.05\pm0.86)\times10^{-3}$   & $-0.94\pm0.53$  & $3.70\pm1.02$      \\%
                       & BaBar \cite{ba06} & $(3.84\pm0.73)\times10^{-3}$   & ~~$1.65\pm0.70$ & 0 (fixed)          \\\hline%
\multirow{2}*{$\Upsilon(4S)\to\Upsilon(2S)\pi^+\pi^-$}
                       & BaBar \cite{ba06} & ~$0.15\pm0.01$                 & $-3.37\pm0.17$  & $11.9\pm3.3$      \\%
                       & BaBar \cite{ba06} & ~$0.14\pm0.01$                 & $-3.67\pm0.14$  & 0 (fixed)          \\
\end{tabular}
\end{ruledtabular}
\end{center}
\end{table}

\begin{figure}[htbp]
\begin{center}\vspace*{0.cm}
\begin{center} 
\hglue0.3cm\includegraphics[width=0.5\textwidth]{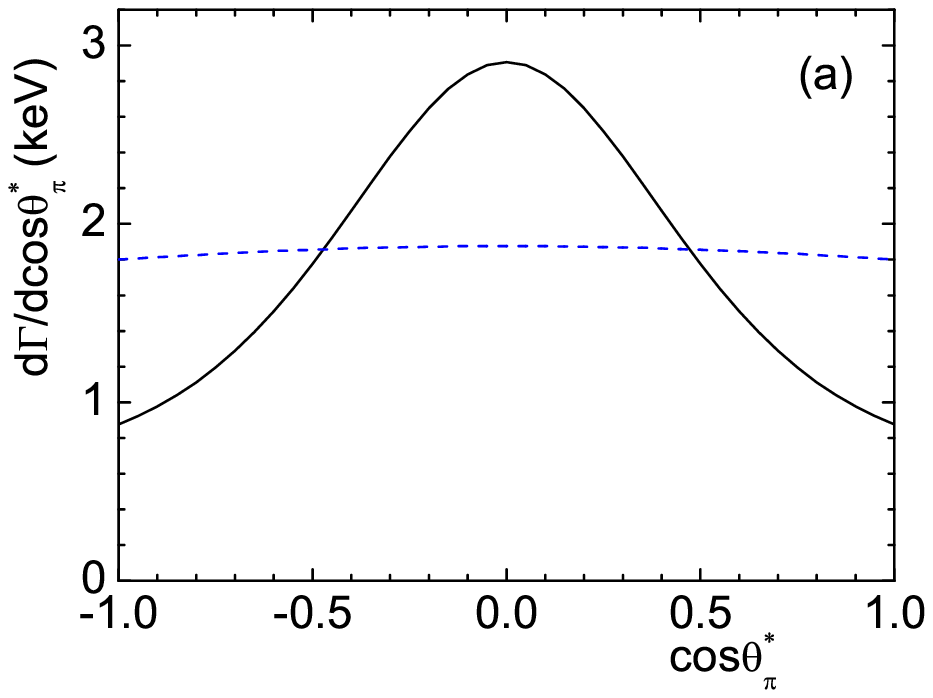}
\hglue-0.6cm\includegraphics[width=0.5\textwidth]{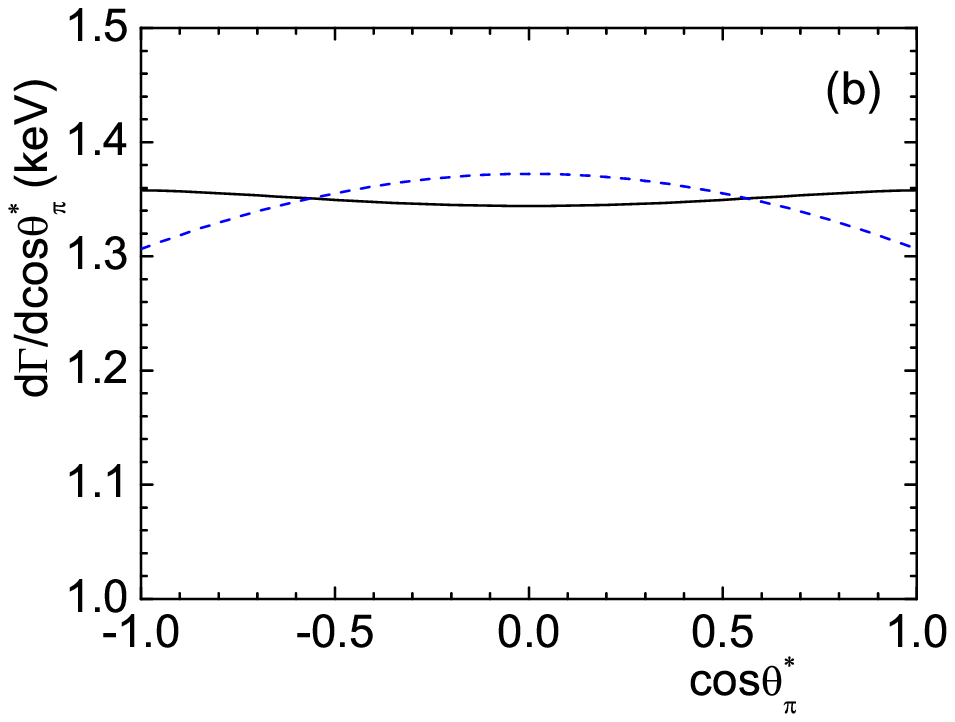}
{\caption{\label{fig:up4Sa} Predictions of the $\cos\theta_{\pi}^*$
distributions. (a) $\Upsilon(4S)\to\Upsilon(1S)\pi^+\pi^-$, (b)
$\Upsilon(4S)\to\Upsilon(2S)\pi^+\pi^-$. The solid and dashed curves
represent the results with and without the intermediate $X$ state,
respectively.}}
\end{center}
\end{center}
\end{figure}

Using the central values of parameters in Table \ref{tab:up4Spara},
we calculate the $\cos\theta_{\pi}^*$ distributions for both
$\Upsilon(4S)$ dipion transitions (for
$\Upsilon(4S)\to\Upsilon(1S)\pi^+\pi^-$, we use the parameter set
determined from the data in Ref. \cite{be07}). The results are
presented in Fig. \ref{fig:up4Sa}, where the solid and dashed curves
represent the results with and without $X$, respectively. From the
figure, one sees that the predicted $\cos\theta_{\pi}^*$
distribution of the $\Upsilon(4S)\to\Upsilon(1S)\pi^+\pi^-$ decay is
much flatter in the case without $X$ than in the case with $X$.
Clearly, the angular distribution of the
$\Upsilon(4S)\to\Upsilon(1S)\pi^+\pi^-$ decay with $X$ exhibits a
similar behavior shown in the
$\Upsilon(3S)\to\Upsilon(1S)\pi^+\pi^-$ decay, because the
sequential process plays dominant role in both decays. Therefore, we
can expect that the future angular distribution data would be a
criterion for further judgment.

Moreover, we would mention that in the
$\Upsilon(4S)\to\Upsilon(1S)\pi^+\pi^-$ decay, the numerical result
shows a negligible effect from the off-shell tensor resonance
$f_2(1270)$.

In summary, we analyze recently reported data of the $\pi^+\pi^-$
invariant mass spectra of the
$\Upsilon(4S)\to\Upsilon(1S,2S)\pi^+\pi^-$ decays in a
model-independent way, and find that the behavior of the $\pi\pi$
invariant mass spectrum in the
$\Upsilon(4S)\to\Upsilon(1S)\pi^+\pi^-$ decay is an analogue of that
in the $\Upsilon(3S)\to\Upsilon(1S)\pi^+\pi^-$ decay, namely there
does exist a broad structure in the $\pi\pi$ mass region below 0.6
GeV. This disagrees with the presumed $\Delta n=2$ rule in Ref.
\cite{vo06}. Then, we try to justify our model developed in studying
the puzzle in the $\Upsilon(3S)\to\Upsilon(1S)\pi^+\pi^-$ decay by
using the newly reported data of $\Upsilon(4S)$ dipion decays. We
find that with the additional contribution from the sequential
process proposed in our model, one can describe the experimental
data, namely the intermediate state $X$ plays an important role in
the $\Upsilon(4S)\to\Upsilon(1S)\pi^+\pi^-$ decay. We also predict
the $\cos\theta_{\pi}^*$ distributions of
$\Upsilon(4S)\to\Upsilon(1S,2S)\pi^+\pi^-$ decays. The result shows
that the $\cos\theta_{\pi}^*$ distribution of the
$\Upsilon(4S)\to\Upsilon(1S)\pi^+\pi^-$ decay in the without $X$
case is much flatter than that in the with $X$ case. Therefore, this
observable can be employed to justify our model in the future.

\begin{acknowledgments}
We would like to thank D. V. Bugg for his constructive comments. We
also benefit much from valuable discussions with B.-S. Zou. We
sincerely appreciate C. Patrignani and A. Sokolov for providing us
the BaBar's and Belle's data, respectively. This work is partially
supported by the NSFC grants under the Nos. 90103020, 10475089,
10435080, 10447130 and CAS Knowledge Innovation Key-Project grant
under the No. KJCX2SWN02.
\end{acknowledgments}

\end{document}